\documentclass[12pt]{article}
\usepackage{epsfig,cite}
\usepackage {amsmath}
\usepackage{amssymb}
\include{epsf}
\newcount\timecount
\newcount\hours \newcount\minutes  \newcount\temp \newcount\pmhours
\hours = \time
\divide\hours by 60
\temp = \hours
\multiply\temp by 60
\minutes = \time
\advance\minutes by -\temp
\def\hour{\the\hours}
\def\minute{\ifnum\minutes<10 0\the\minutes
            \else\the\minutes\fi}
\def\clock{
\ifnum\hours=0 12:\minute\ AM
\else\ifnum\hours<12 \hour:\minute\ AM
      \else\ifnum\hours=12 12:\minute\ PM
            \else\ifnum\hours>12
                 \pmhours=\hours
                 \advance\pmhours by -12
                 \the\pmhours:\minute\ PM
                 \fi
            \fi
      \fi
\fi
}

\def\monthname{\relax\ifcase\month 0/\or January\or February\or
   March\or April\or May\or June\or July\or August\or September\or
   October\or November\or December\else\number\month/\fi}

\def\bold#1{\setbox0=\hbox{$#1$}%
     \kern-.025em\copy0\kern-\wd0
     \kern.05em\copy0\kern-\wd0
     \kern-.025em\raise.0433em\box0 }

\def\beq{\begin{equation}}
\def\eeq{\end{equation}}

\def\ss{\scriptscriptstyle}
\def\ga{\mathrel{\raise.3ex\hbox{$>$\kern-.75em\lower1ex\hbox{$\sim$}}}}
\def\la{\mathrel{\raise.3ex\hbox{$<$\kern-.75em\lower1ex\hbox{$\sim$}}}}
\def\gev{{\rm \, Ge\kern-0.125em V}}
\def\tev{{\rm \, Te\kern-0.125em V}}
\def\gyr{{\rm \, G\kern-0.125em yr}}

\def\gappeq{\mathrel{\rlap {\raise.5ex\hbox{$>$}}
{\lower.5ex\hbox{$\sim$}}}}
\def\lappeq{\mathrel{\rlap{\raise.5ex\hbox{$<$}}
{\lower.5ex\hbox{$\sim$}}}}
\def\Toprel#1\over#2{\mathrel{\mathop{#2}\limits^{#1}}}

\def\m12{m_{1\!/2}}

\def\mz{m_{\ss Z}}

\def\PL{{Phys.~Lett.} }

\def\NP{{Nucl. Phys.} }

\def\tanb{\tan \beta}

\def\swsq{\sin^2 \theta_W}

\def\cthsq{\cos^2 \theta_{\tilde t}}
\def\sthsq{\sin^2 \theta_{\tilde t}}

\def\bea{\begin{eqnarray}}
\def\eea{\end{eqnarray}}
\newcommand{\goto}{\rightarrow}
\newcommand{\bmm}{B_s \goto \mu^+ \, \mu^-}
\newcommand{\mbs}{M_{B_s}}
\newcommand{\fbs}{f_{B_s}}
\newcommand{\mchr}[1]{m_{\chi^+_{ #1}}}
\newcommand{\msup}[1]{m_{\tilde{u}_{ #1}}}

\begin{document}
\begin{titlepage}
\pagestyle{empty}
\baselineskip=21pt
\rightline{\tt hep-ph/0504196}
\rightline{CERN-PH-TH/2005-062}
\rightline{UMN--TH--2352/05}
\rightline{FTPI--MINN--05/10}
\vskip 0.2in
\begin{center}
{\large{\bf On the Interpretation of {\boldmath $B_s \to \mu^+ 
\mu^-$} in the CMSSM}}
\end{center}
\begin{center}
\vskip 0.2in
{\bf John~Ellis}$^1$, {\bf Keith~A.~Olive}$^{2}$
and {\bf Vassilis~C.~Spanos}$^{2}$
\vskip 0.1in
{\it
$^1${TH Division, CERN, Geneva, Switzerland}\\
$^2${William I. Fine Theoretical Physics Institute, \\
University of Minnesota, Minneapolis, MN 55455, USA}}

\vskip 0.2in
{\bf Abstract}
\end{center}
\baselineskip=18pt \noindent

We discuss the interpretation of present and possible future experimental
constraints on $\bmm$ decay in the context of the constrained minimal
extension of the Standard Model (CMSSM) with universal scalar masses. We
emphasize the importance of including theoretical and other experimental
uncertainties in calculating the likelihood function, which can affect
significantly the inferred 95 \% confidence-level limit on the CMSSM
parameters. The principal uncertainties are the $B_s$ meson decay
constant, $m_t$ and $m_b$. The latter induce uncertainties in the mass of
the neutral Higgs boson that dominates the $\bmm$ decay amplitude at large
$\tan \beta$, reducing the CMSSM region excluded by present and possible
future limits from the Fermilab Tevatron collider and the LHC.

\vfill
\leftline{CERN-PH-TH/2005-062}
\leftline{July 2005}
\end{titlepage}
\baselineskip=18pt

\section{Introduction}

The prospects for new physics searches at the LHC and other future
colliders are already constrained by rare processes that are sensitive to
small deviations from the Standard Model. A prime example of such a
low-energy constraint is $b \to s \gamma$ decay \cite{bsg,bsgth}. This, the anomalous
magnetic moment of the muon, $g_\mu - 2$ \cite{g-2}, and the Higgs mass, $m_h$ \cite{mh},
are among the most important indirect constraints on extensions of the
Standard Model, such as the minimal supersymmetric extension of the
Standard Model (MSSM).

The decay $B_s \to \mu^+ \mu^-$ has been emerging as another interesting
potential constraint on the parameter space of models for physics beyond
the Standard Model, such as the MSSM \cite{Dedes,Arnowitt,ko,baer}. 
The Fermilab Tevatron collider
already has an interesting upper limit on the branching ratio for $B_s \to
\mu^+ \mu^-$ \cite{cdf}, and is expected soon to increase significantly its
sensitivity to $B_s \to \mu^+ \mu^-$ decay, whilst the LHC sensitivity
will reach down to the Standard Model rate \cite{Buras}. Since its branching ratio is
known to increase rapidly for large values of the ratio of Higgs v.e.v.'s,
$\tan \beta$ \cite{calcs,Babu,Bobeth}, increasing like its sixth power, these present and future
sensitivities are particularly important for models with large $\tan
\beta$.

In view of its potential importance for the MSSM, it is important to treat
the $B_s \to \mu^+ \mu^-$ decay constraint with some care, as has already
been done for the $b \to s \gamma$, $g_\mu - 2$ and $m_h$ constraints. In
each of these cases, the interpretation depends on uncertainties in
theoretical calculations, which should be propagated carefully and
combined with the experimental errors if limits are to be calculated at
some well-defined confidence level, or if a global fit to MSSM parameters
is to be attempted.

Both of these issues are important also in the case of $B_s \to \mu^+
\mu^-$ decay. As concerns the theoretical uncertainties, it is important
to include consistently all the available one-loop MSSM contributions, and
avoid any approximate treatments of any individual terms, since
cancellations are potentially important, and also to include knowledge of
the higher-order QCD corrections to the lowest-order MSSM loop diagrams.  
As we discuss in this paper, other sources of error and uncertainty are
also important. These include, for example, the uncertainties in the $B_s$
meson parameters, principally the decay constant $f_{B_s}$. Since the $B_s
\to \mu^+ \mu^-$ decay rate is dominated by the exchange of the
pseudoscalar Higgs boson $A$, the value of $m_A$ is also an important
uncertainty.

Bounds on supersymmetry are often explored in a constrained model with
universal soft supersymmetry-breaking scalar masses $m_0$, gaugino masses
$m_{1/2}$ and trilinear couplings $A_0$ at some GUT input scale. In this
CMSSM, the values of $m_A$ and magnitude of the Higgs mixing parameter
$\mu$ (but not its sign) are in principle fixed by the electroweak
vacuum conditions. However, these predictions are necessarily approximate.
For example, the value of $m_A$ predicted as a function of the independent
parameters $m_{1/2}, m_0, A_0$ and $\tan \beta$ has significant
uncertainties associated with the lack of precision with which the heavy
quark masses $m_t$ and $m_b$ are known, as we discuss extensively later in
this paper. Moreover, rather different values of $m_A$ would be predicted
in models where the universality assumptions of the CMSSM are relaxed. For
example, much smaller values of $m_A$ are attainable in models with
non-universal Higgs masses (NUHM).

When interpreting experimental upper bounds (or measurements) within any
specific model such as the CMSSM, care must be taken to incorporate the
uncertainties in auxiliary parameters such as $f_{B_s}$, $m_t$ and $m_b$.  
These should be propagated and combined with the experimental likelihood
function when quoting sensitivities in, e.g., the $(m_{1/2}, m_0)$ plane
at some fixed level of confidence. Moreover, one must also be aware of
model dependences within the assumed framework such as the value of $A_0$
in the CMSSM, as well as the effects of possible deviations from the model
framework such as non-universal Higgs masses.

We exemplify these points in a discussion of uncertainties in the
interpretation of the present and likely future sensitivities of the
Fermilab Tevatron collider and the LHC to $B_s \to \mu^+ \mu^-$ decay,
assuming $\mu > 0$ as preferred by $g_\mu - 2$. In particular, we show
that the uncertainties in $f_{B_s}$, $m_t$ and $m_b$ each shrink
significantly the regions that might otherwise appear to be excluded by
the present limit, or might appear to be if the decay is not discovered at
the likely future sensitivity. We compare the resulting $B_s \to \mu^+
\mu^-$ constraints with other existing constraints such as $b \to s
\gamma$, discussing how they vary with $A_0$ and commenting on the 
situation within the NUHM.

\section{Calculation of {\boldmath $\bmm$} Decay}

The branching ratio for the decay $\bmm$ is given by
\bea
\mathcal{B}(\bmm) &=& \frac{G_F^2 \alpha^2}{16 \pi^3} \frac{\mbs^5 \fbs^2  
\tau_B }{4} |V_{tb}V_{ts}^*|^2 \sqrt{1-\frac{4 m_\mu^2}{\mbs^2}} \nonumber \\
  &\times& \left\{    \left(1-\frac{4 m_\mu^2}{\mbs^2}\right) | C_S |^2 
  + \left |C_P-2 \, C_A \frac{m_\mu}{\mbs^2} \right |^2   \right\} \, ,
\label{eq:braratio}
\eea
where the one-loop corrected Wilson coefficients $C_{S,P}$ are taken 
from~\cite{Bobeth} and $C_A$ is defined in terms of $Y(x_t)$, 
following~\cite{Logan}, as $C_A=Y(x_t)/\swsq$ where
\beq
Y(x_t)=1.033 \left( \frac{m_t(m_t)}{170 \gev} \right)^{1.55} \, .
\eeq
The function $Y(x_t)$ incorporates both leading \cite{Inami} and
next-to-leading order~\cite{Buras} QCD corrections, and $m_t(m_t)$ is the
running top-quark mass in the $\overline{MS}$ scheme. 
The precise value of $m_t(m_t)$ depends somewhat
on the set of supersymmetric parameters and our choice of the physical top
quark mass $m_t=178 \pm 4$~GeV~\cite{D0} that we use in this paper.
The Wilson coefficients $C_{S,P}$ have been multiplied by
$1/(1+\epsilon_b)^2$, where $\epsilon_b$ incorporates the full
supersymmetric one-loop correction to the bottom-quark Yukawa
coupling~\cite{mbcor,Carena, Pierce}. It is known that, since $\epsilon_b$ is
proportional to $\tanb$, this correction may be significant in the
large-$\tanb$ regime we study here \cite{Dedes,Arnowitt}.
These corrections to $\epsilon_b$ are flavour independent.
In addition, it is important to include the flavour-violating 
contributions arising from the Higgs and chargino couplings
at the one-loop level. These corrections result in effective
one-loop corrected values for the  Kobayashi-Maskawa (KM)
matrix elements~\cite{Babu,Isidori}, which we include as described 
in~\cite{Buras1,Tata}. In particular, these corrections modify the Wilson
coefficients involved in Eq.~(\ref{eq:braratio}), as can be seen 
in Eqs. (6.35) and (6.36) in ~\cite{Buras1} or in Eq. (14) in ~\cite{Tata}.
In addition, the latter study includes the effects of squark mixing, which are included here as well.
Below, we discuss the behaviour of the dominant 
contribution to the Wilson coefficients, mainly  to illustrate their
dependence on the pseudo-scalar Higgs boson mass.

The Wilson coefficients $C_{S,P}$ receive four contributions in the
context of MSSM, due to Higgs bosons, counter-terms, box and penguin
diagrams.  The Higgs-boson corrections were calculated in~\cite{Logan},
and the rest of the supersymmetric corrections in~\cite{calcs,Babu}.  
The full one-loop corrections have been studied and presented
comprehensively in~\cite{Bobeth}. 
In our numerical analysis, we implement the full
one-loop corrections taken from this work 
as well as the dominant NLO QCD corrections
studied in~\cite{Buras2}, as well as
the flavour-changing gluino contribution~\cite{Tata,Bobeth2} .
The Higgs-boson, box and
penguin corrections to $C_{S,P}$ are proportional to $\tan^2\beta$, while
the counter-term corrections dominate in the large-$\tan \beta$ limit, as
they are proportional to $\tan^3\beta$. 

In order to understand the behaviour of the branching ratio in the
$(m_{1/2},m_0)$ plane in the context of the CMSSM, we focus attention on
the counter-terms which are mediated by $A,H,h$ exchange as 
seen in Eqs. (5.1) and (5.2) of~\cite{Bobeth}. As seen in Eq. (5.13) of~\cite{Bobeth},
the $\bmm$ decay amplitude $\propto 1/m_{A}^2$ in the large-$\tan \beta$
limit, and the decay rate $\propto 1/m_{A}^4$.  This underlines the
importance of knowing or calculating $m_{A}$ as accurately as
possible. 

The counter-term contribution to $C_{S,P}$ is given by \cite{Bobeth}
\bea\label{susy:result:count}
C_{S,P}^{\rm CT} &=&\mp
\frac{m_\mu\tan^3\beta}{\sqrt{2}M_W^2
m_{A}^2}\sum_{i=1}^{2}\sum_{a=1}^{6}\sum_{m,n=1}^{3}
 [m_{\tilde{\chi}_i^{\pm}} D_3(y_{ai})U_{i2}(\Gamma^{U_L})_{am}\Gamma^a_{imn}],
\eea
where 
\beq\label{susy:result:gamma}
\Gamma^a_{imn}= \frac{1}{2\sqrt{2}\sin^2\theta_W}
[\sqrt{2}M_W V_{i1}(\Gamma^{U_L\dagger})_{na}-(M_U)_{nn}V_{i2}
     (\Gamma^{U_R\dagger})_{na}]\lambda_{mn},
\eeq
and $M_U \equiv  {\rm diag}(m_u, m_c, m_t)$. $U$ and $V$ are the chargino
diagonalization matrices, $\Gamma^{U_L}$ and  $\Gamma^{U_R}$ are $6 \times 3$
squark diagonalization matrices, and $D_3(x) \equiv x\ln x/(1-x)$. 
Additionally, $y_{ai}$ is defined in
Eq. (5.10) of~\cite{Bobeth} as $y_{ai} \equiv \msup{a}^2/\mchr{i}^2$, 
where $\msup{a}^2 \equiv \{ m_{{\tilde{u}_L}}^2, m_{{\tilde{c}_L}}^2, 
m_{{\tilde{t}_1}}^2,
m_{{\tilde{u}_R}}^2, m_{{\tilde{c}_R}}^2,  m_{{\tilde{t}_2}}^2 \}$. 
Finally, $\lambda_{mn} \equiv V_{mb}V^*_{ns}/V_{tb}V^*_{ts}$.

We can split the counter-term contribution  into two terms:  
one that is proportional to $M_W$ and another that is
proportional to $m_t$.
Starting with the term whose numerator depends on $M_W$,  
it is easy to see that the non-vanishing terms  
stem from the following combinations of 
indices:
$\{a,n,m\}=\{111,222,333,633\}$ and $i=1,2$. However, the first term 
$\{111\}$ is
suppressed, as it is  proportional to
$\lambda_{11}=V_{ub}\, V_{us}/V_{tb}\, V_{ts} \simeq - 0.022$, 
whereas the 
second is not suppressed, because it is
proportional to
$\lambda_{22}=V_{cb}\, V_{cs}/V_{tb}\, V_{ts} \simeq - 1$. 
Nor are the third and fourth terms suppressed, as they are multiplied by  
$\lambda_{33}=1$.
Thus, the part of the counter-term contribution to the Wilson coefficient 
that is $\propto M_W$ is 
\bea
C_S^{CT,M_W}=&-& \sqrt{2} M_W \, f  \left\{  \mchr{1}\, V_{11}\,U_{12} 
\left[ \lambda_{22}\, D_3(y_{21})+ \lambda_{33}\, \left( \cthsq \, D_3(y_{31}) + \sthsq \, 
D_3(y_{61} ) \right)  \right] \right. \nonumber \\
&+& \left.   \mchr{2}\, V_{21}\,U_{22} 
\left[ \lambda_{22}\,D_3(y_{22})+ \lambda_{33}\, \left(\cthsq \, D_3(y_{32}) + \sthsq 
\, D_3(y_{62}) \right) \right]    \right\} ,
\label{eq:ctmw}
\eea
where
\beq
f \equiv \frac{m_\mu \tan^3\beta }{4M_W^2 \swsq m_A^2} \, ,
\label{eq:ffact}
\eeq
and we have ignored in (\ref{eq:ctmw}) terms that are proportional 
to $\lambda_{11}$. The unitarity 
of the KM 
matrix implies that $\lambda_{11}+\lambda_{22}+\lambda_{33}=0$, which
for small $\lambda_{11}$ means $\lambda_{22}=-\lambda_{33}$, resulting
in the suppression of $C_S^{CT,M_W}$.

Turning now to the terms that increase with the charge-2/3 quark masses,
we see that the terms with $n=3$ (the top-quark contributions) dominate
the first- and second-generation terms in $\Gamma_{imn}^a$.  
Specifically, the dominant terms have $\{a,n,m\}=\{333,633\}$. In
addition, we notice that the $i=2$ part is important, since it is
proportional to $V_{22} \, U_{22}$, while the $i=1$ term is proportional
to $V_{12} \, U_{12}$. Hence it is sufficient to take
\beq
C_S^{CT,m_t} = m_t  \, f \,\, \left(  \frac{\sin2\theta_{\tilde t}}{2}\right)
                     \mchr{2}\,      [  D(y_{32})-D(y_{62}) ] \, ,
\label{eq:ctmt}
\eeq
where we set $V_{22}\simeq U_{22} \simeq 1$. 
The expression 
(\ref{eq:ctmt}) is the approximation derived in~\cite{Arnowitt}. 
The GIM cancellation in  (\ref{eq:ctmw}) means that the counterterm
contribution to the Wilson coefficients, which is the dominant one, can be 
approximated relatively well by (\ref{eq:ctmt}). However,
in our analysis we use the full expressions given in~\cite{Bobeth}, as
well as the gluino correction and the flavour-violating corrections to
Kobayashi-Maskawa matrix elements as described above.

\section{Dependence of {\boldmath $m_A$}  on {\boldmath $m_t$} and on 
{\boldmath $m_b$}}

It is clear from the discussion above that the mass of the
pseudoscalar Higgs boson, $m_{A}$, is an important ingredient in
calculating the branching ratio for the decay $\bmm$, since it enters in
the fourth power in the denominators of the Wilson coefficients $C_S$ and
$C_P$ in (\ref{eq:braratio}). Therefore, to further our discussion of the
uncertainties in the $\bmm$ branching ratio, we now discuss the
uncertainties in the calculated value of $m_A$.

The electroweak symmetry breaking conditions may be written in the form:
\begin{equation}
m_A^2  = m_{H_1}^2 + m_{H_2}^2 + 2 \mu^2 + \Delta_A
\label{eq:mamass}
\end{equation}
and
\begin{equation}
\mu^2 = \frac{m_{H_1}^2 - m_{H_2}^2 \tan^2 \beta + \frac{1}{2} \mz^2 (1 - \tan^2 \beta)
+ \Delta_\mu^{(1)}}{\tan^2 \beta - 1 + \Delta_\mu^{(2)}},
\label{eq:minmu}
\end{equation}
where $\Delta_A$ and $\Delta_\mu^{(1,2)}$ are loop
corrections~\cite{Pierce,Barger:1993gh,deBoer:1994he,Carena:2001fw,erz}.
The exact forms of the radiative corrections to $\mu$ and $m_A$
are not needed here, but it is important to note that, at large $\tan 
\beta$, the dominant contribution to $\Delta_\mu^{(1)}$ contains a term
which is proportional to $h_t^2 \tan \beta^2$, whereas the dominant
contribution to $m_A^2$ contains terms proportional to 
$h_t^2 \tan \beta$ and $h_b^2 \tan \beta$.  
Therefore, the $m_{H_2}^2$ term along with a piece proportional 
to $h_t^2$ in $\Delta_\mu^{(1)}$ are dominant in (\ref{eq:minmu}) in the 
large-$\tanb$ regime, so $\mu$ depends rather mildly on $m_b$.

\begin{figure}
\vskip 0.5in
\vspace*{-0.75in}
\begin{minipage}{8in}
\epsfig{file=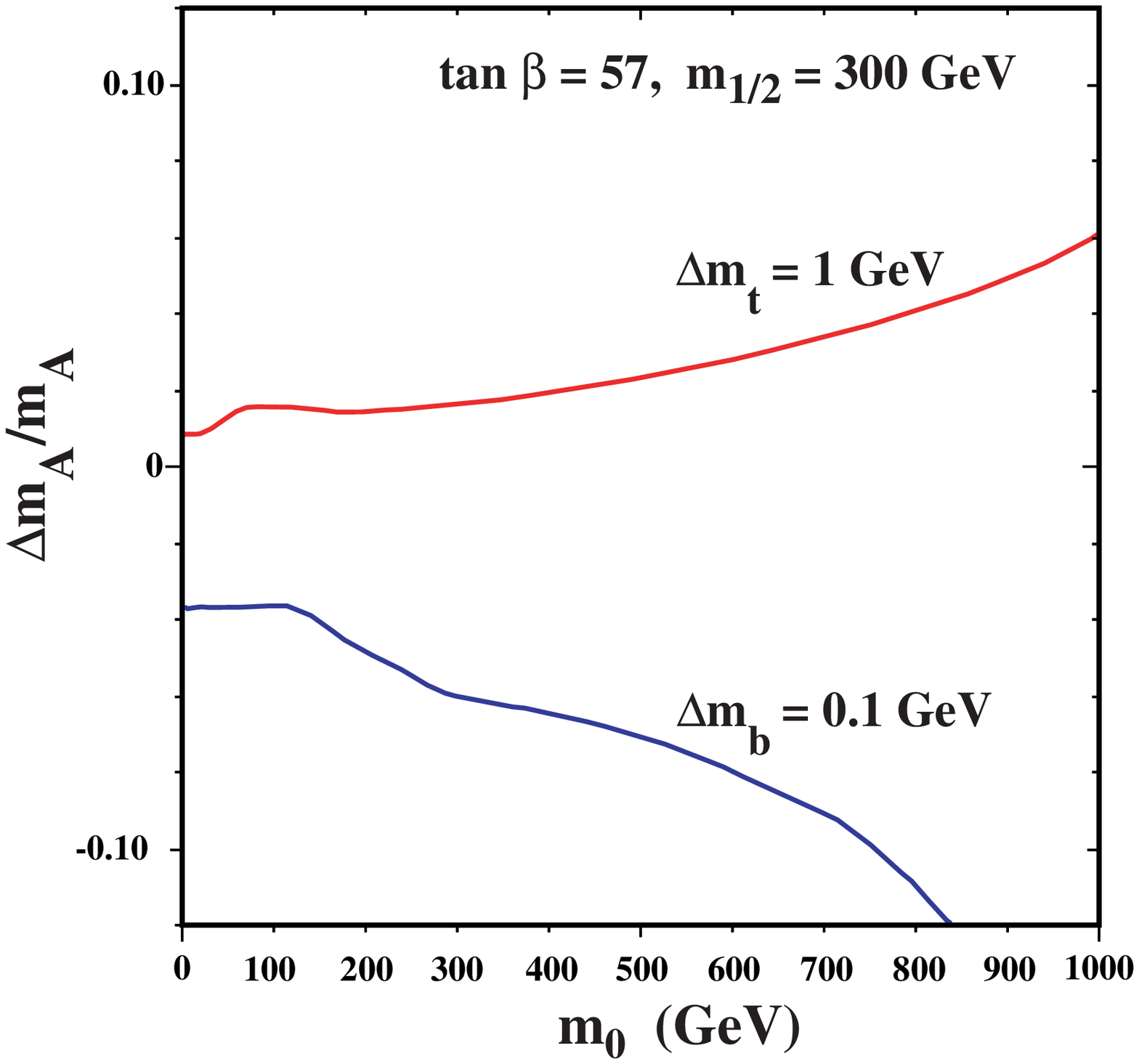,height=3in}
\hspace*{-0.17in}
\epsfig{file=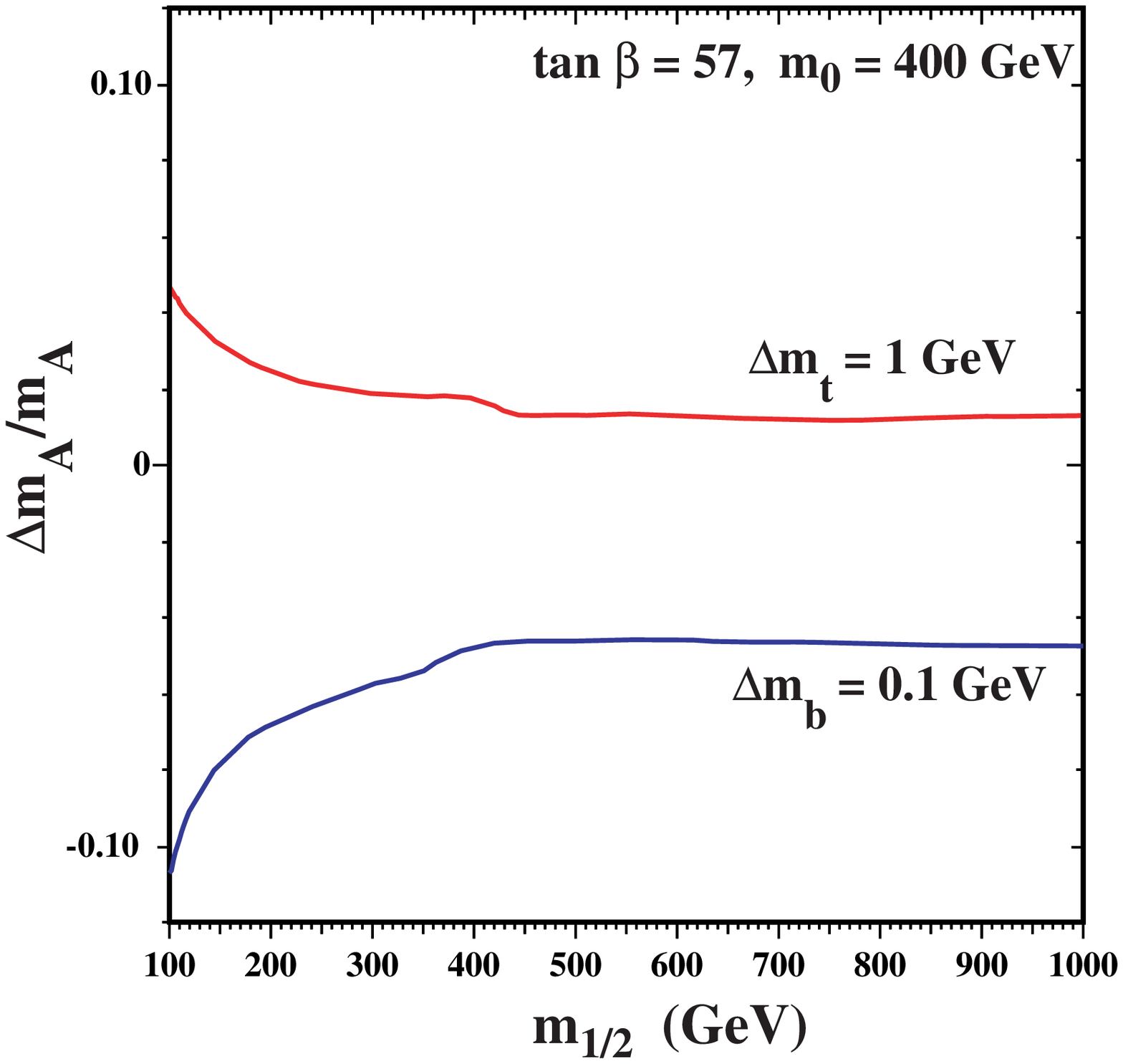,height=3in}
\hfill
\end{minipage}
\caption{
{\it
The sensitivities of $m_A$ to $m_t$ and $m_b$, assuming $\Delta m_t = 
1$~GeV 
and $\Delta m_b = 0.1$~GeV, along slices through the $(m_{1/2}, m_0)$ 
CMSSM plane for $A_0 = 0$ and $\tan \beta = 57$. Panel (a) fixes $m_{1/2} 
= 300$~GeV and varies $m_0$, while panel (b) fixes $m_0
= 400$~GeV and varies $m_{1/2}$.}
} 
\label{fig:errors} \end{figure}

We illustrate in Fig.~\ref{fig:errors} the logarithmic sensitivity of 
$m_A$, namely $\Delta m_A/m_A$, to $m_b$ and $m_t$ along slices through 
the $(m_{1/2}, m_0)$ plane for $\tan \beta = 57$, $A_0 = 0$ and $\mu > 0$. 
We use as representative errors $\Delta m_t = 1$~GeV and $\Delta 
m_b = 0.1$~GeV. 
Panel (a) shows the effect of varying $m_0$ for fixed $m_{1/2} = 
300$~GeV, and panel (b) shows the effect of varying $m_{1/2}$ for fixed 
$m_0 = 400$~GeV. 

One can understand the behaviour depicted by employing (\ref{eq:mamass}),
(\ref{eq:minmu}) and the renormalization-group equations (RGEs) that
govern the evolution of the $m_{H_i}$ from $M_{GUT}$ to $M_Z$. The
one-loop RGE for $m_{H_1}^2$ depends on $h_b$:
\beq
\frac{\partial m_{H_1}^2}{\partial \ln Q}=\frac{1}{16 \pi^2} 
\left\{  (-6g_2^2 \, M_2^2-\frac{6}{3}g_1^2 \,M_1^2)\,
+\,6|h_b|^2(m_{Q_3}^2 +  m_{D_3}^2 + m_{H_1}^2 +A_b^2 )     \right\} \, ,
\label{eq:rgeh1}
\eeq
where we ignore, for simplicity, the $h_\tau$ contribution, while the RGE 
for $m_{H_2}^2$ depends on $h_t$:
\beq
\frac{\partial m_{H_2}^2}{\partial \ln Q}=\frac{1}{16 \pi^2} 
\left\{  (-6g_2^2 \, M_2^2-\frac{6}{3}g_1^2 \,M_1^2)\,
+\,6|h_t|^2(m_{Q_3}^2 +  m_{U_3}^2 + m_{H_2}^2 +A_t^2 )     \right\} \, .
\label{eq:rgeh2}
\eeq

We see from (\ref{eq:rgeh1}) that, as $m_b$ increases, the value of
$m_{H_1}^2$ at $M_Z$ decreases, tending to decrease $m_A$
(\ref{eq:mamass}). More importantly, the term proportional to $h_b^2$ in
$\Delta_A$ in (\ref{eq:mamass}) is proportional to $m_{{\tilde
b}_1}^2/(m_{{\tilde b}_1}^2 - m_{{\tilde b}_2}^2)\log(m_{{\tilde b}_1}^2/
m_{{\tilde b}_2}^2) - 1$, which is negative across the $(m_{1/2}, m_0)$
plane for $\tan \beta = 57$ and $A_0 = 0$, with a magnitude that decreases
with $m_{1/2}$. Thus, as $m_b$ increases, we obtain a further decrease in
$m_A$. This term in fact provides most of the numerical dependence of
$m_A$ on $m_b$. Since it is enhanced by $\tan\beta$, the $m_b$ dependence
becomes milder for smaller values of $\tan\beta$.

In addition, the effect on $m_A$ is augmented if the role of the bottom
Yukawa coupling in the RGE (\ref{eq:rgeh1}) is enhanced, which occurs at 
large $m_0$. This increases the sensitivity
of $m_A$ to $m_b$, as seen in Fig. \ref{fig:errors}(a). In contrast,
when $m_{1/2}$ is increased, the sensitivity to $\Delta_A$ is diminished
and, at the same time, the gaugino part of the RGE is enhanced. Both
effects lead to a reduced change in $m_A$ at large $m_{1/2}$, as can be
seen in Fig. \ref{fig:errors}(b).

Turning now to the dependence of $m_A$ on $m_t$, we see that the evolution
of $m_{H_2}^2$ shown in (\ref{eq:rgeh2}) is similar to (\ref{eq:rgeh1}),
apart from the substitution of $h_t$ and analogous mass substitutions. As
$m_t$ increases, the value of $m_{H_2}^2$ at $M_Z$ is driven to larger
negative values. However, the change in $m_A$ is dominated by the change
in $\mu$, which grows with an increase in $m_t$. The net effect is an
increase in $m_A$, as seen in Fig. \ref{fig:errors}, which increases with 
$m_0$ and decreases with $m_{1/2}$, as seen in panels (a) and (b), 
respectively.

Since the $\bmm$ decay rate depends on the fourth power of $m_A$, the
sensitivity of $m_A$ to both $m_b$ and $m_t$ displayed in
Fig.~\ref{fig:errors} can lead to a large uncertainty in $\bmm$ for $\tan
\beta = 57$. We have also evaluated the sensitivities to $m_t$ and $m_b$
for $\tan \beta = 40$. These sensitivities do not vary significantly with
$m_{1/2}$ nor with $m_0$. They are always smaller
than for $\tan \beta = 57$, and the difference is rather substantial for
large $m_0$ and small $m_{1/2}$.

Numerically, in the following analysis we assume
\begin{equation}
\Delta m_t \, = \, 4~{\rm GeV}, \; \Delta m_b \, = \, 0.11~{\rm GeV},
\label{deltamtmb}
\end{equation}
with central values for the physical pole mass $m_t = 178$~GeV and the 
running mass $m_b^{\overline {MS}}(m_b) = 4.25$~GeV.
The first of the uncertainties in (\ref{deltamtmb}) is taken directly from 
measurements at 
the Fermilab Tevatron collider, and may be reduced by a factor of 2 to 4 
by future measurements there and at the LHC. The following analysis shows 
that such reductions would be most welcome also in the analysis of $\bmm$ 
decay. The uncertainty in $m_b$ is taken from a recent review~\cite{E-KL} 
that combines determinations from ${\bar b}b$ systems, $b$-flavoured 
hadrons and high-energy processes. Our 1-$\sigma$ range given 
in~\cite{E-KL} is contained 
within the preferred range quoted by the Particle Data Group~\cite{pdg}, 
and is very 
similar to the ranges quoted recently by the UKQCD group~\cite{UKQCD} 
in the unquenched approximation and 
in the review 
given by Rakow at the Lattice 2004 conference~\cite{Rakow}.

It is easy to see how important these uncertainties could be. For example,
when $\Delta m_A/m_A = 0.05$ for $\Delta m_t = 1$~GeV, which occurs when
$\tan \beta = 57$ for $(m_{1/2}, m_0)  = $ (300, 900) or (100, 400)~GeV,
the change in $m_A$ for $|\Delta m_t| = 4$~GeV is $\pm 0.2 m_A$,
corresponding to a change in the $A$ contribution to the $\bmm$ decay rate
by a factor 2.07 or 0.41, depending on the sign of $\Delta m_t$. 

We note in passing that both the CDF~\cite{CDFH} and D0~\cite{D0H}
collaborations have recently published new upper limits on Higgs
production at the Fermilab Tevatron collider. In particular, the D0
limit~\cite{D0H} is relevant to the MSSM at very large $\tan \beta$ and
small $m_A$. However, such small values of $m_A$ are already excluded at
large $\tan \beta$ in the CMSSM by other constraints such as $b \to s
\gamma$ and the lower limit on $m_h$, so the present D0 limit does not
further restrict the part of the CMSSM parameter space of interest here.

\section{The Effects of Uncertainties on {\boldmath $\bmm$} Limits in the 
CMSSM}

\begin{figure}
\vskip 0.5in
\vspace*{-0.75in}
\begin{minipage}{8in}
\epsfig{file=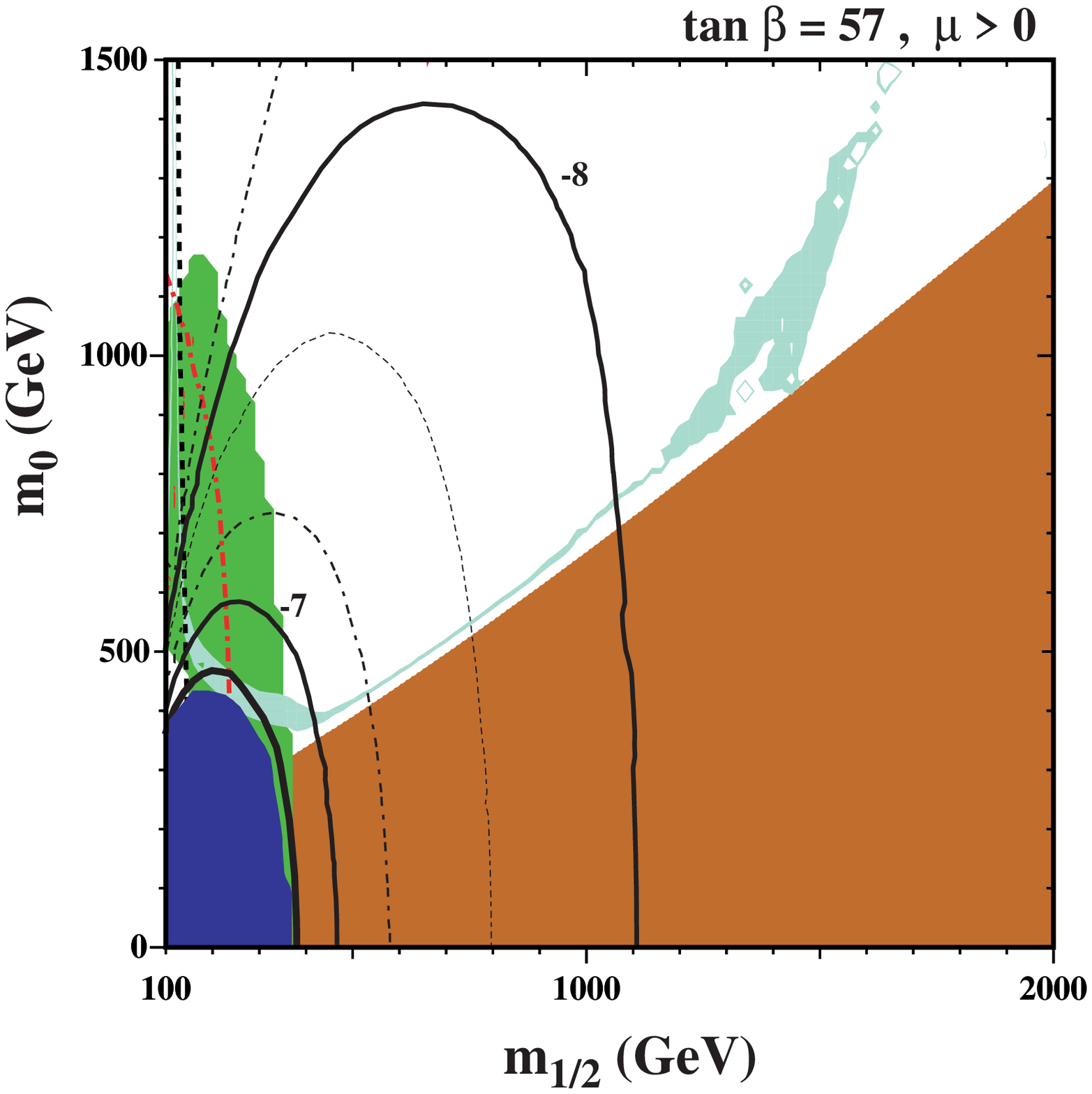,height=3.3in}
\hspace*{-0.17in}
\epsfig{file=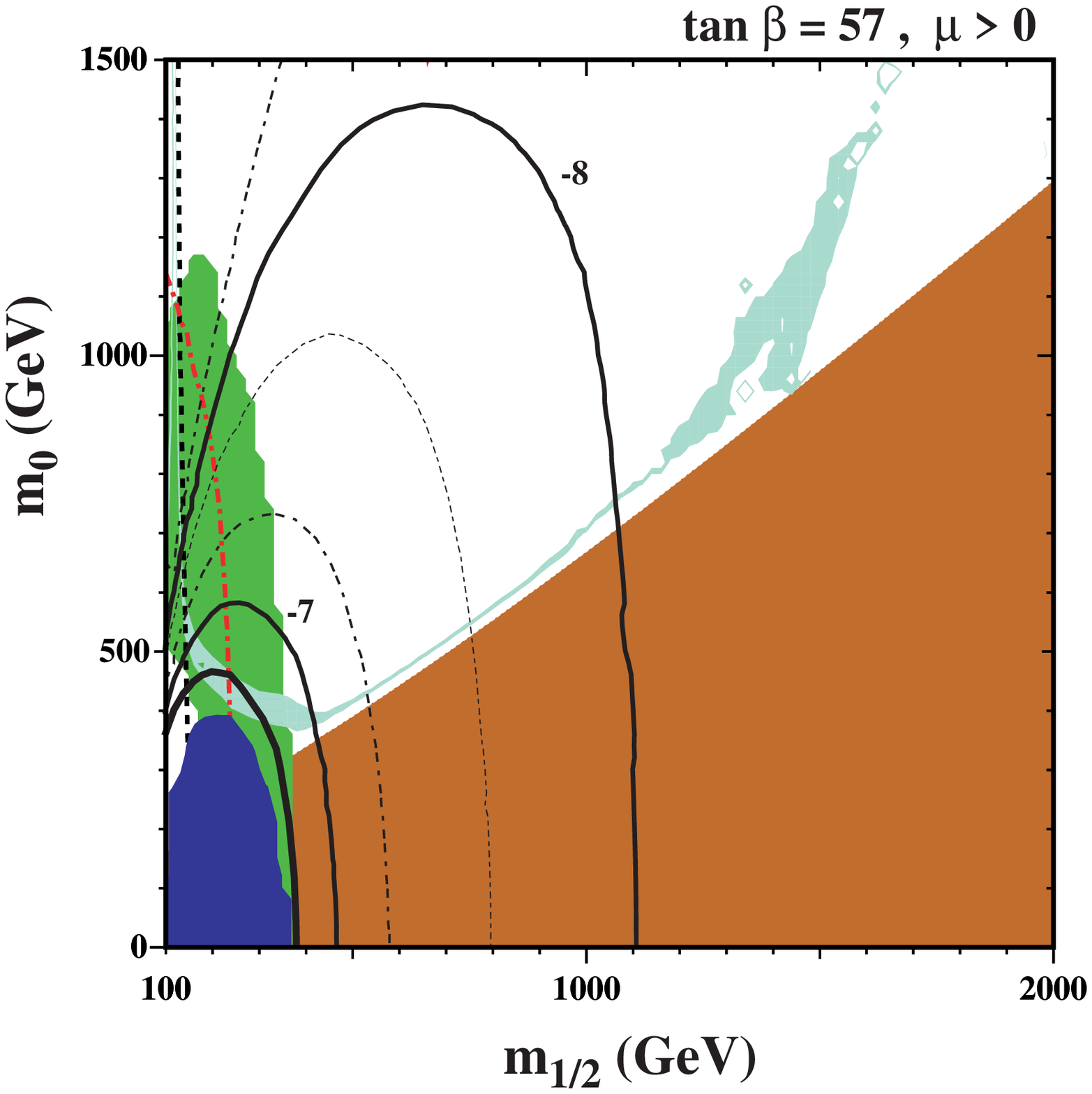,height=3.3in}
\hfill
\end{minipage}
\begin{minipage}{8in}
\epsfig{file=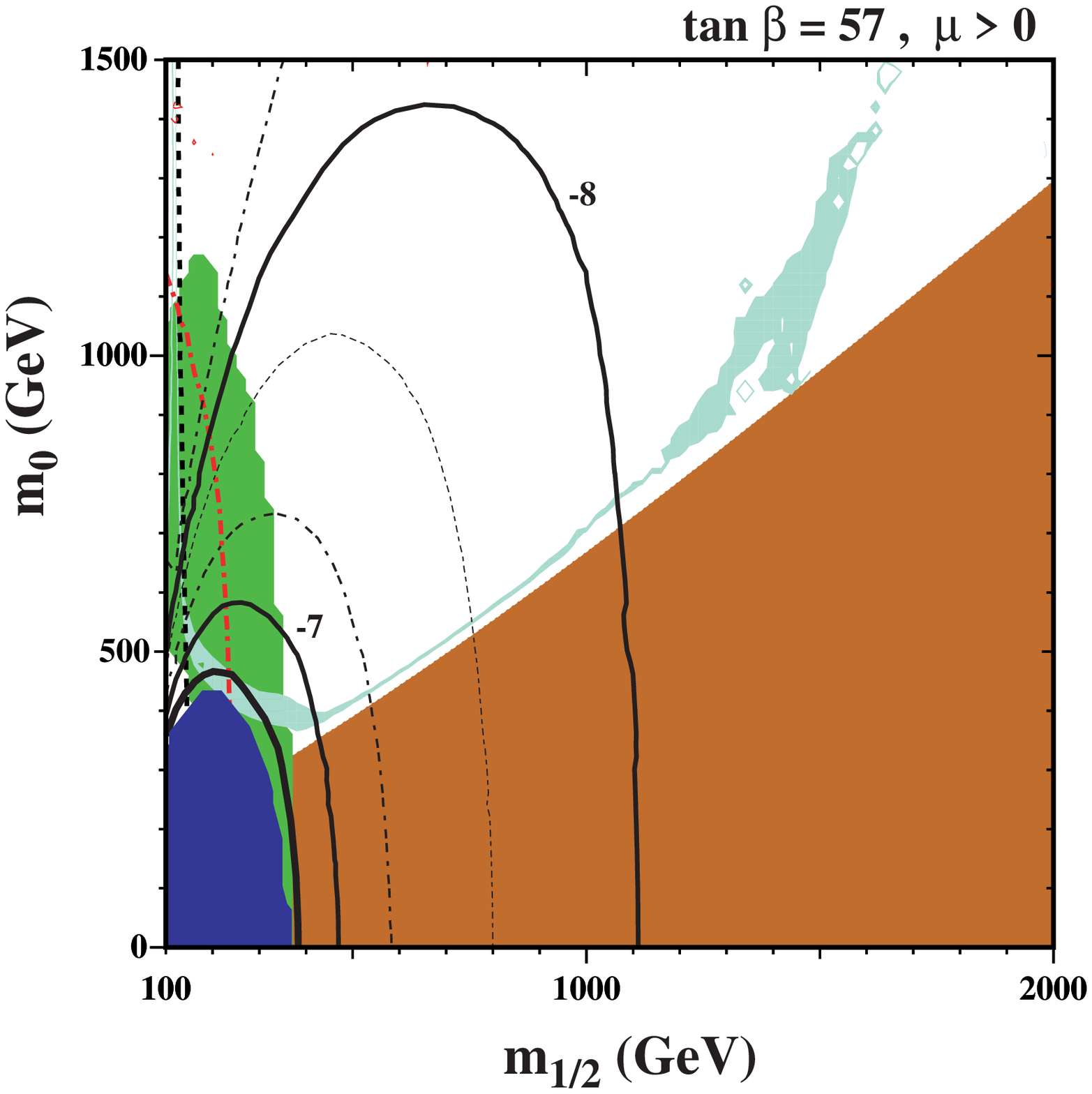,height=3.3in}
\hspace*{-0.2in}
\epsfig{file=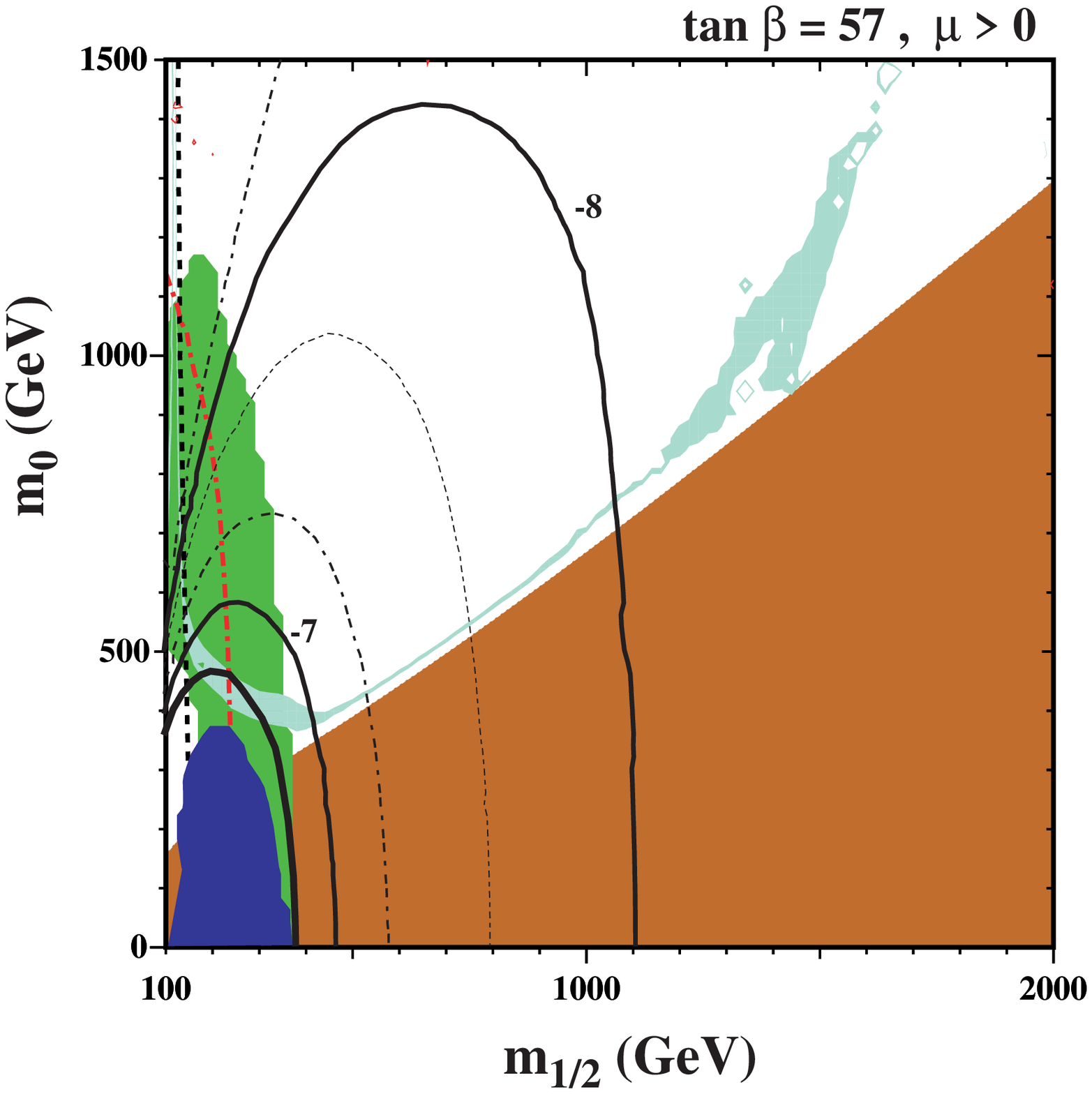,height=3.3in} \hfill
\end{minipage}
\caption{
{\it
The effects of auxiliary uncertainties on the region of the $(m_{1/2}, 
m_0)$ plane for $A_0 = 0$, $\mu > 0$ and $\tan \beta = 57$ currently 
excluded by the 
Fermilab Tevatron collider. (a) The effect of $B_s$ meson uncertainties 
alone, principally that in $f_{B_s}$. (b) These uncertainties combined 
with the uncertainty $\Delta m_t = 4$~GeV. (c) The $B_s$ meson 
uncertainties combined with the uncertainty $\Delta m_b = 0.11$~GeV. (d) 
All uncertainties combined. The various contours and shadings in the 
$(m_{1/2}, m_0)$ plane are explained in the text.}
} 
\label{fig:CMSSMerrors} 
\end{figure}

In order to assess how important these auxiliary uncertainties may be in
the interpretation of $\bmm$ experiments, we display in
Fig.~\ref{fig:CMSSMerrors} their individual effects on the present $\bmm$
constraint in the $(m_{1/2}, m_0)$ plane of the CMSSM for $A_0 = 0$, $\mu
> 0$ and $\tan \beta = 57$. This is close to the largest value of $\tan
\beta$ for which we find suitable electroweak vacua in generic domains of
the $(m_{1/2}, m_0)$ plane, and so maximizes the potential impact of the
$\bmm$ constraint, which increases asymptotically as the sixth power of
$\tan \beta$. The dark (brick) shaded regions in the bottom-right corners
of each panel are excluded because there the lightest supersymmetric
particle (LSP) would be the charged ${\tilde \tau_1}$. The pale (blue)  
shaded strips are those favoured by WMAP, if all the cold dark matter is
composed of LSPs. The supersymmetric spectrum 
and relic density calculations have been descibed elsewhere
(see e.g. \cite{else}).  The near-vertical dashed (black) lines at small
$m_{1/2}$ are the constraint imposed by the non-observation of charginos
at LEP, and the near-vertical dash-dotted (red) lines are those imposed by
the non-observation of the lightest MSSM Higgs boson, as calculated using
the {\tt FeynHiggs} code \cite{FeynHiggs}. 

The medium (green) shaded regions are excluded
by the rare decay $b \to s \gamma$. 
The branching ratio for this  has been measured
by the CLEO, BELLE and BaBar collaborations~\cite{bsg}. 
The theoretical prediction of $b \to s 
\gamma$~\cite{bsgth,Ambrosio,Buras1} contains uncertainties which stem from 
the uncertainties
in $m_b$, $\alpha_s$, the measurement of the semileptonic branching ratio
of the $B$ meson, and the effect of the scale dependence. 
In particular, the scale dependence of the theoretical prediction arises from
the dependence on three scales: the scale where the QCD corrections to the
semileptonic decay are calculated and the high and low energy scales
relevant to $b \to s \gamma$ decay~\cite{scale}. 
These sources of uncertainty can be
combined to determine a total 
theoretical uncertainty~\footnote{According to a recent 
analysis~\cite{Neubert},
these theoretical uncertainties may be significantly larger, resulting to
a weaker bound on the masses of supersymmetric particles.}.
The experimental measurement is converted into a Gaussian likelihood and
convolved with a theoretical likelihood to determine the total 
likelihood~\cite{efgo},
which is used to calculate the excluded region at 95\% CL.
It is important to note that the dependence of this excluded 
region on $m_A, m_b,$ and $m_t$ is quite weak in comparison, 
as we have checked numerically. 

Finally, the ellipsoidal contours represent the nominal $\bmm$ branching
ratio, calculated (like all the others) using the current central values
$m_t = 178$~GeV and $m_b^{\overline {MS}}(m_b) = 4.25$~GeV. The numerical
labels for the two outer solid lines are exponents in the branching ratio:
$10^{-7}, 10^{-8}$, the thinner-dashed line is for $2 \times 10^{-8}$, the
thicker-dashed line for $5 \times 10^{-8}$.

The most stringent experimental upper bound on the $\bmm$ branching 
ratio is that given by an updated CDF measurement:
$1.5 \times 10^{-7}$ ($2.0 \times 10^{-7})$ at the 90\% (95\%) 
CL~\cite{cdf}.
The innermost thick solid line of  Fig.~\ref{fig:CMSSMerrors} is
the contour for the present nominal 95\% CL experimental upper limit of 
$2.0 \times
10^{-7}$.
 
Panel (a) of Fig.~\ref{fig:CMSSMerrors} displays as a blue shaded region
the effect on the interpretation of the present experimental limit of
including the uncertainties in the $B_s$ meson parameters. The most
important uncertainty is that in the decay constant $f_{B_s}$, for which
we assume~\cite{bernard}
\beq
f_{B_s} = 230 \pm 30~{\rm MeV}. 
\eeq
In addition, we use~\cite{pdg}
\beq
m_{B_s}=5369.6 \pm 2.4~{\rm MeV}, \qquad \tau_{B_s}=(1.461 \pm 0.057) \times
10^{-12}~{\rm s}.
\eeq
In calculating the uncertainty we add
quadratically the uncertainties that result from these errors as well as
those in the KM elements. We see that incorporating these
uncertainties does not change the overall shape of the excluded region,
but does shrink it slightly. There may be some possibility to reduce
the uncertainty in $f_{B_s}$ in the foreseeable future, but in this
analysis we retain it fixed in the following panels and other figures.

The blue shaded region in panel (b) of Fig.~\ref{fig:CMSSMerrors}
incorporates the present uncertainty in $m_t$, assumed to be $\Delta m_t =
4$~GeV, which is propagated through the CMSSM calculation of $m_A$ as
discussed in the previous Section. We see that this uncertainty is more
important for larger $m_0$, truncating the upper part of the exclusion
domain. This effect can readily be understood from panel (a) of
Fig.~\ref{fig:errors}, where we saw that $\Delta m_t$ has a particularly
important effect on $m_A$ at large $m_0$.

The blue shaded region in panel (c) of Fig.~\ref{fig:CMSSMerrors} shows
the parallel effect of the uncertainty in $m_b$, assumed to be $\Delta m_b
= 0.11$~GeV, which is also propagated through the CMSSM calculation of
$m_A$ as discussed in the previous Section. This uncertainty is also more
important for larger $m_0$, providing an independent mechanism for
truncating the upper part of the exclusion domain. This can also readily
be understood from panel (a) of Fig.~\ref{fig:errors}, where we saw that
$\Delta m_b$ also has a particularly important effect on $m_A$ at large
$m_0$.

There is also some tendency in both panels (b) and (c) of
Fig.~\ref{fig:CMSSMerrors} for the exclusion domain to separate from the
axis $m_{1/2} \sim 100$ GeV, particularly at large $m_0$. This can be understood
from panel (b) of Fig.~\ref{fig:errors}, where we see that the effects of
both $\Delta m_t$ and $\Delta m_b$ on $m_A$ are enhanced when $m_{1/2} \la
200$~GeV. The uncertainties in each of $m_t$ and $m_b$ become
particularly important when $m_{1/2}$ is small and $m_0$ large, as seen
separately in panels (b) and (c) of Fig.~\ref{fig:CMSSMerrors}.

The similar tendencies in panels (b) and (c) of Fig.~\ref{fig:CMSSMerrors}
are reinforced when we combine the uncertainties in $m_t$ and $m_b$, as
shown by the blue shaded region in panel (d). We find that $\bmm$ decay
is currently unable to exclude any value of $m_0$ above about 350~GeV,
whereas the exclusion region would have extended up to $m_0 \sim 450$~GeV
if the auxiliary uncertainties had not been taken into account, and $\sim
400$~GeV if either of the $m_t$ or $m_b$ uncertainties had been ignored.  
On the other hand, the reduction in the excluded range of $m_{1/2}$ at
lower $m_0$ is less important, typically $\la 30$~GeV.

We observe that the region currently excluded by $\bmm$ is always included
within the region already excluded by $b \to s \gamma$ and/or $m_h$, even
without including the auxiliary uncertainties. The same is even more true
for smaller values of $\tan \beta$: in the case $\tan \beta = 40$ (not
shown), the region currently excluded by $\bmm$ has $m_{1/2} \la 180$~GeV
and $m_0 \la 170$~GeV, within the strips excluded by $m_h$ and
$m_{\chi^\pm}$ but allowed by $b \to s \gamma$.  We recall that the $b \to
s \gamma$ limit is very dependent on the assumption of universal scalar
masses for the squarks, which does not play a role elsewhere in the
analysis of constraints on CMSSM parameters, and is of course untested.
Clearly $\bmm$ has the potential to complement or even, in the future,
supplant the $b \to s \gamma$ constraint, though it also relies on
squark-mass universality.

\begin{figure}
\vskip 0.5in
\vspace*{-0.75in}
\begin{minipage}{8in}
\epsfig{file=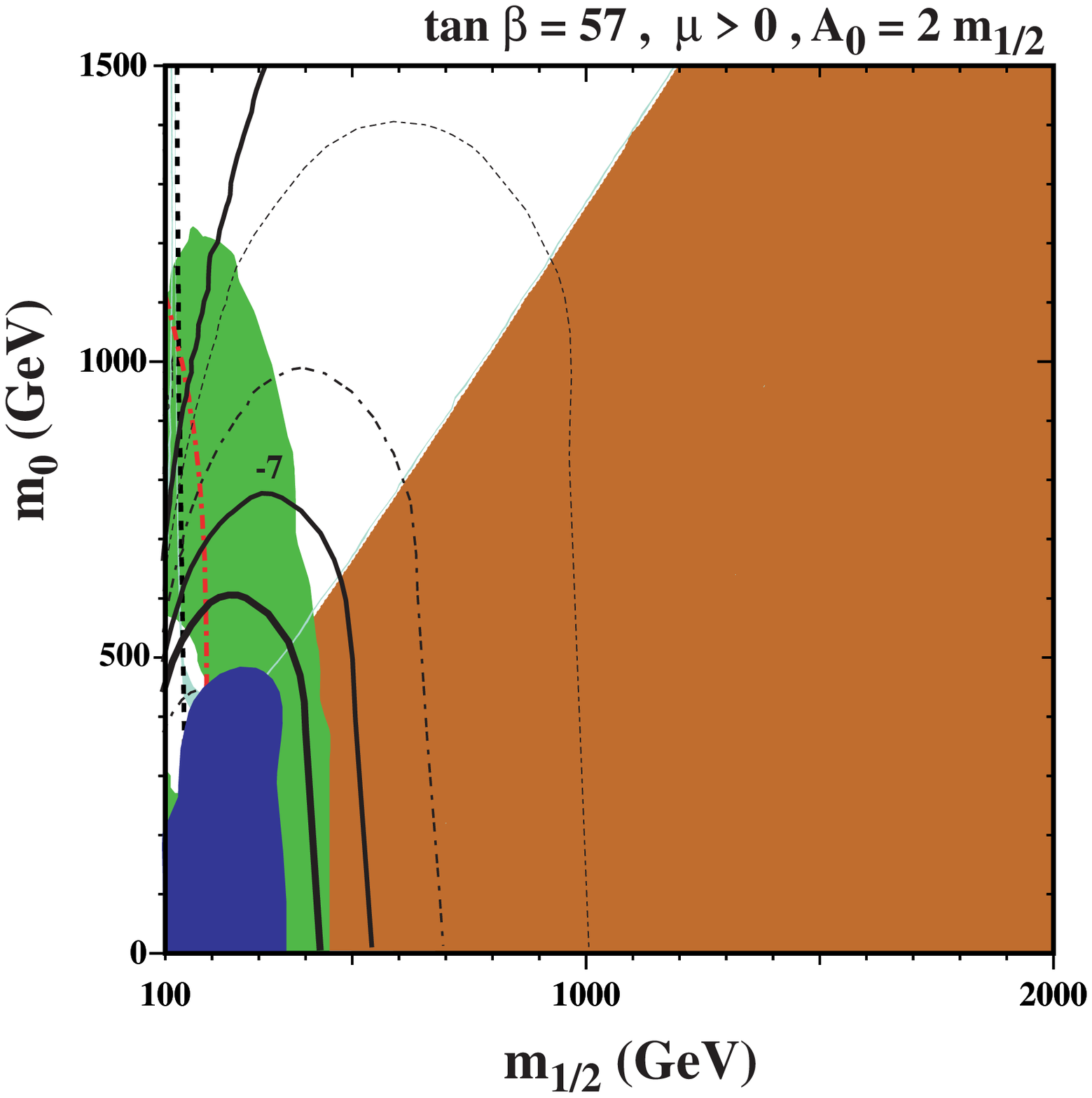,height=3.3in}
\hspace*{-0.17in}
\epsfig{file=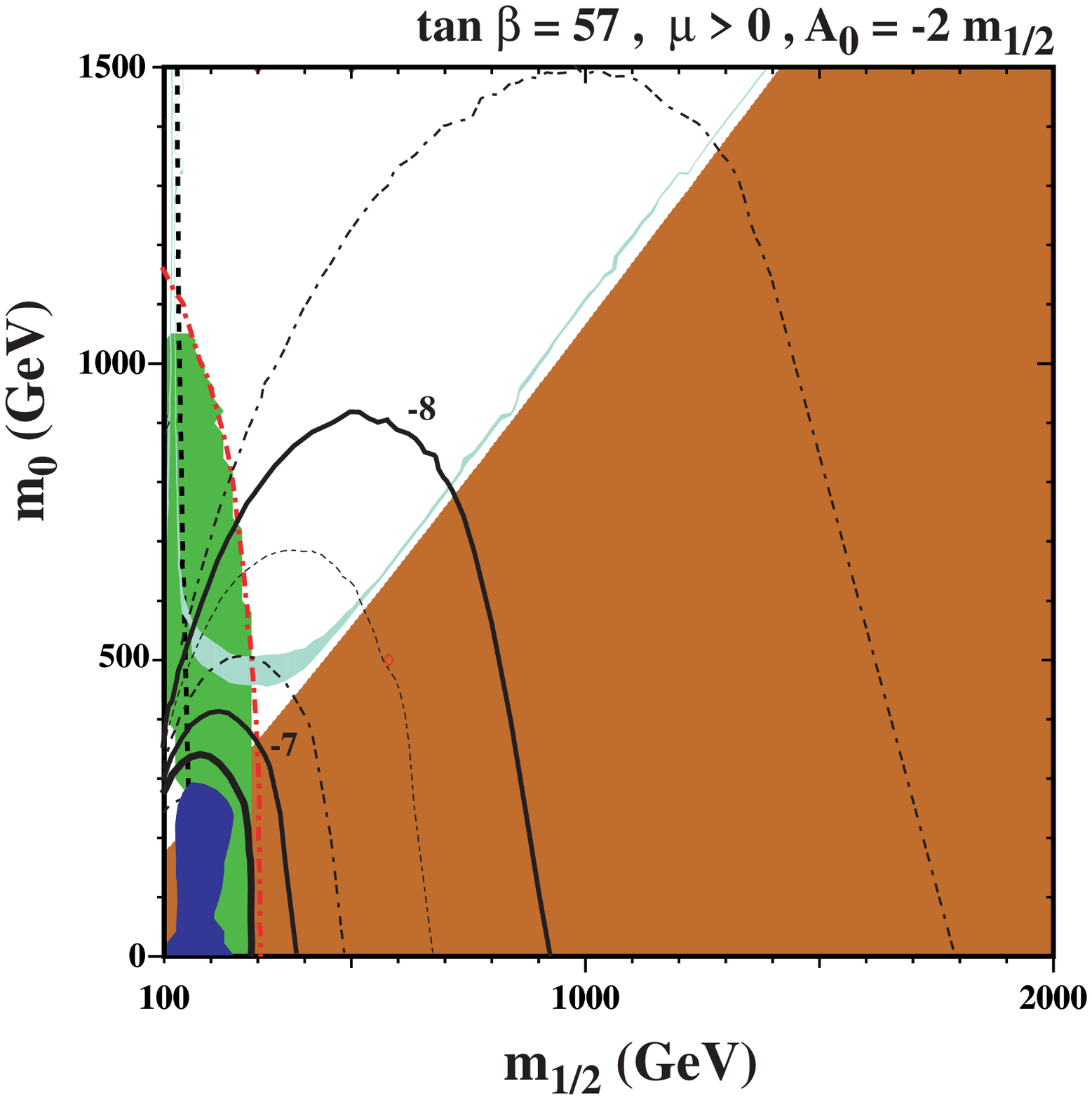,height=3.3in}
\hfill
\end{minipage}
\caption{
{\it
The disallowed regions in the $(m_{1/2}, m_0)$ plane for 
$\mu > 0$ and $\tan \beta = 57$, for (a) $A_0 = 2 m_{1/2}$ and (b) $A_0 = 
- 2 m_{1/2}$. The various contours and shadings in the $(m_{1/2}, m_0)$ 
plane are as explained in the text describing Fig.~\ref{fig:CMSSMerrors}.
}}
\label{fig:varyA}
\end{figure}

The input value of the trilinear soft supersymmetry-breaking parameter
$A_0$ has a significant effect on the allowed CMSSM parameter space, and
is also important for $m_h$ as well as the $b \to s \gamma$ and $\bmm$
decays. Therefore, we display in Fig.~\ref{fig:varyA} the interplays of
these constraints for $\tan \beta = 57$, $\mu > 0$, and (a) $A_0 = 2
m_{1/2}$ and (b) $A_0 = - 2 m_{1/2}$. The qualitative conclusions are
similar to the $A_0 = 0$ case discussed previously: the region currently
disallowed by $\bmm$ decay largely overlaps with the regions previously
disfavoured by $m_h$ and $b \to s \gamma$, and decreases in extent as 
$A_0$ is reduced from positive to negative values.

\section{Treatment of Errors}

There are several ways to treat the auxiliary errors in the $\bmm$ 
analysis. In the above, we have implicitly assumed one of the most 
conservative treatments, in the sense that it excludes the smallest region 
of the $(m_{1/2}, m_0)$ plane for given fixed values of the uncertainties. 
However, other treatments are possible, and here we compare their results. 
For this comparison, we include all the uncertainties in $f_{B_s}$, $m_t$ 
and $m_b$ discussed in the previous Section.

In our previous treatment, see panel (d) of Fig.~\ref{fig:CMSSMerrors}, we
assumed that all the uncertainties have Gaussian error distributions, and
defined the allowed region by discarding the upper 2.5\% tail of the
likelihood distribution obtained by combining the experimental and
auxiliary errors.  This would be correct if the central value of the
experimental measurement, after subtracting any backgrounds, was strictly
zero. Alternatively, one might discard the upper 5 \% of the combined
likelihood distribution.
This would give the correct experimental upper limit if the central
experimental value were far enough above zero that no significant part of
the lower tail of the likelihood distribution extends below zero $\bmm$
decay rate, but is otherwise clearly more conservative than the previous
treatment.  Finally, experimentalists sometimes subtract one theoretical
(systematic) error from the measured result and then plot the 95 \%
confidence-level contour given by the experimental error.

The two alternative prescriptions
yield similar upper limits on $m_0$, though the shapes of the allowed regions
are different. They both reach up to $m_0 \sim 400$~GeV,
as compared the range $m_0 \la 350$~GeV found in the previous analysis,
shown in panel (d) of Fig.~\ref{fig:CMSSMerrors}. We prefer to use the
more conservative prescription used in drawing the previous figures, also
because it is demonstrably appropriate if the central experimental value
is negligible, as is currently the case.

\section{Possible Future Developments}

\begin{figure}
\vskip 0.5in
\vspace*{-0.75in}
\begin{minipage}{8in}
\epsfig{file=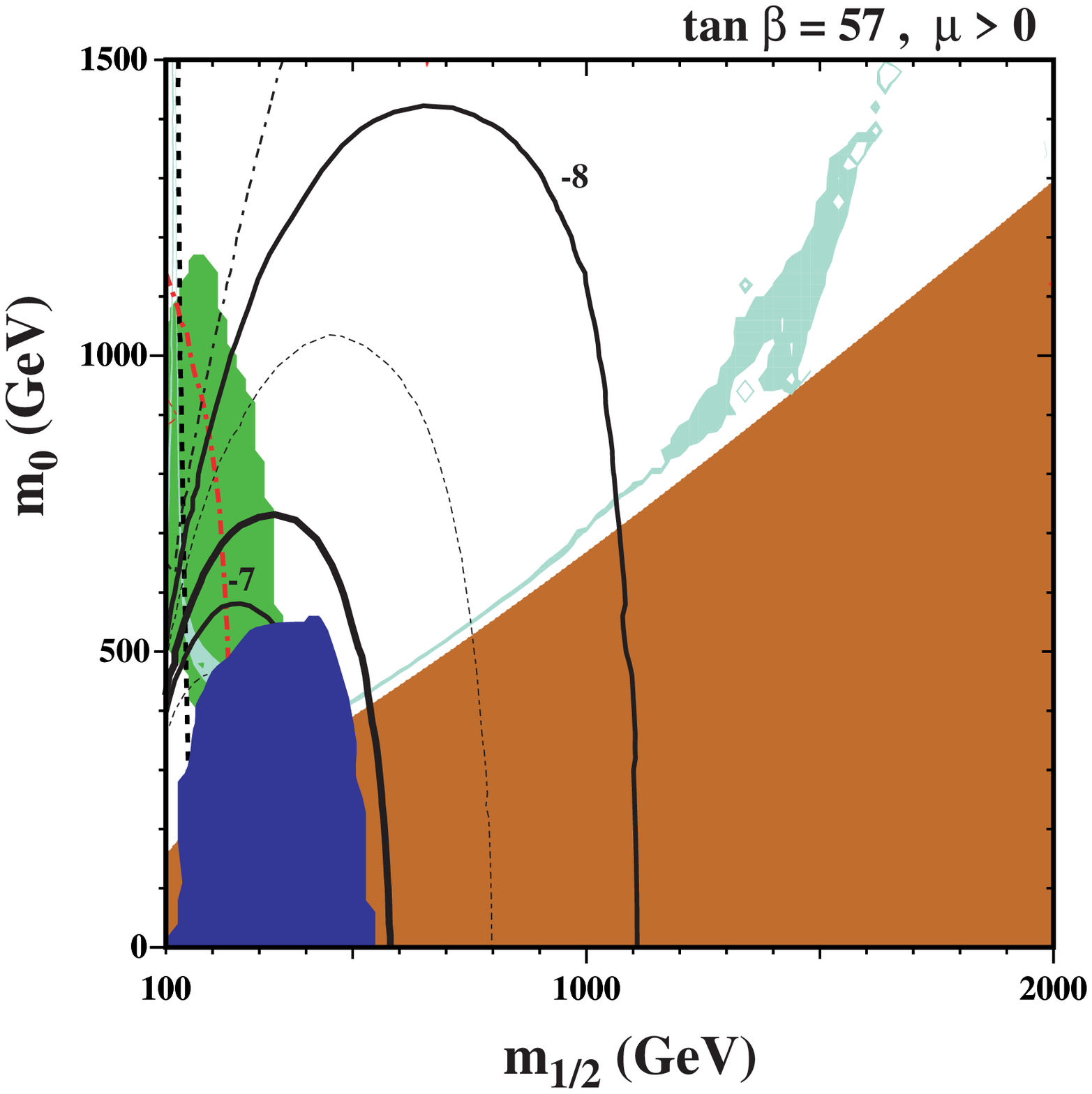,height=3.3in}
\hspace*{-0.17in}
\epsfig{file=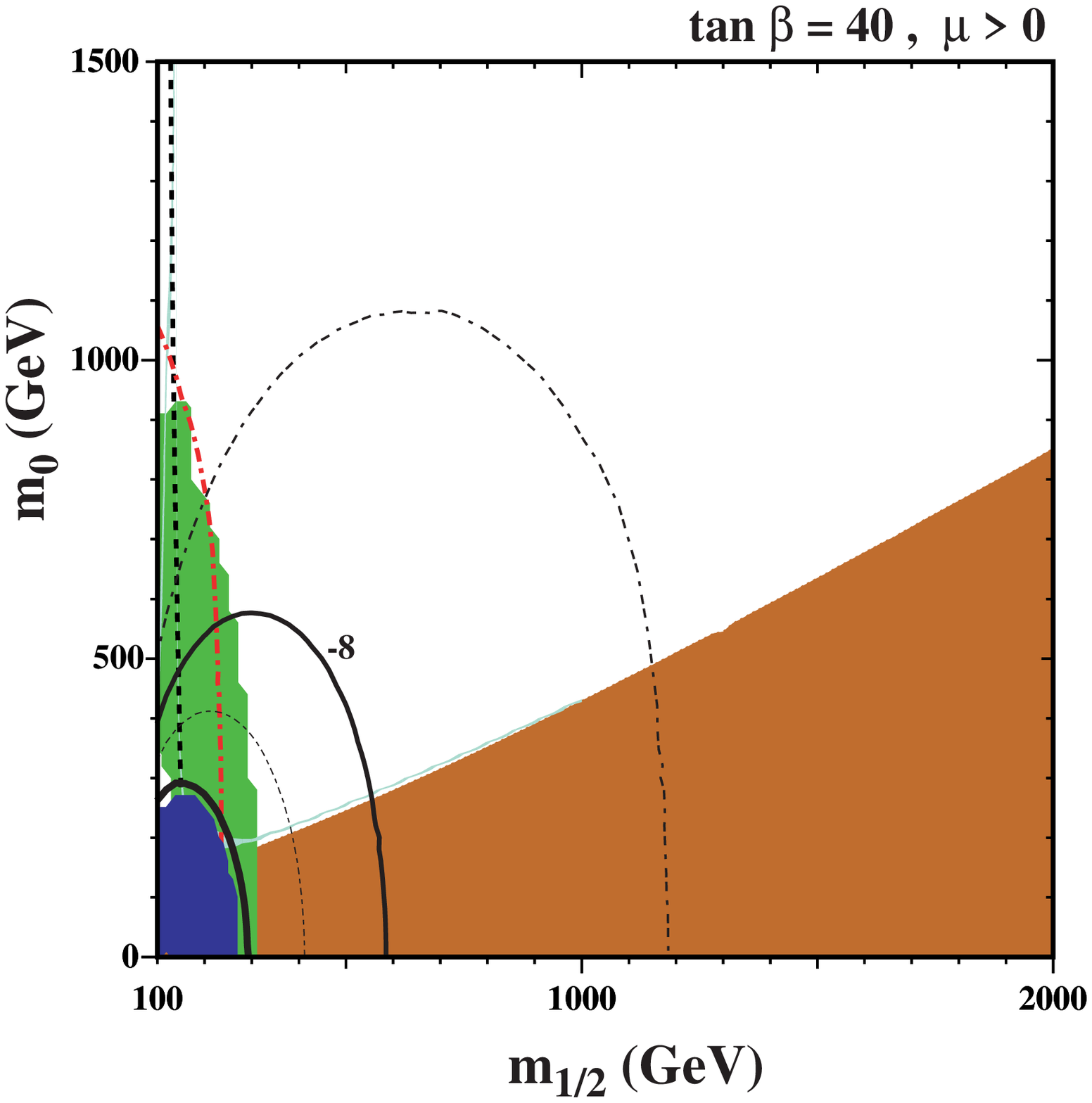,height=3.3in}
\hfill
\end{minipage}
\begin{minipage}{8in}
\epsfig{file=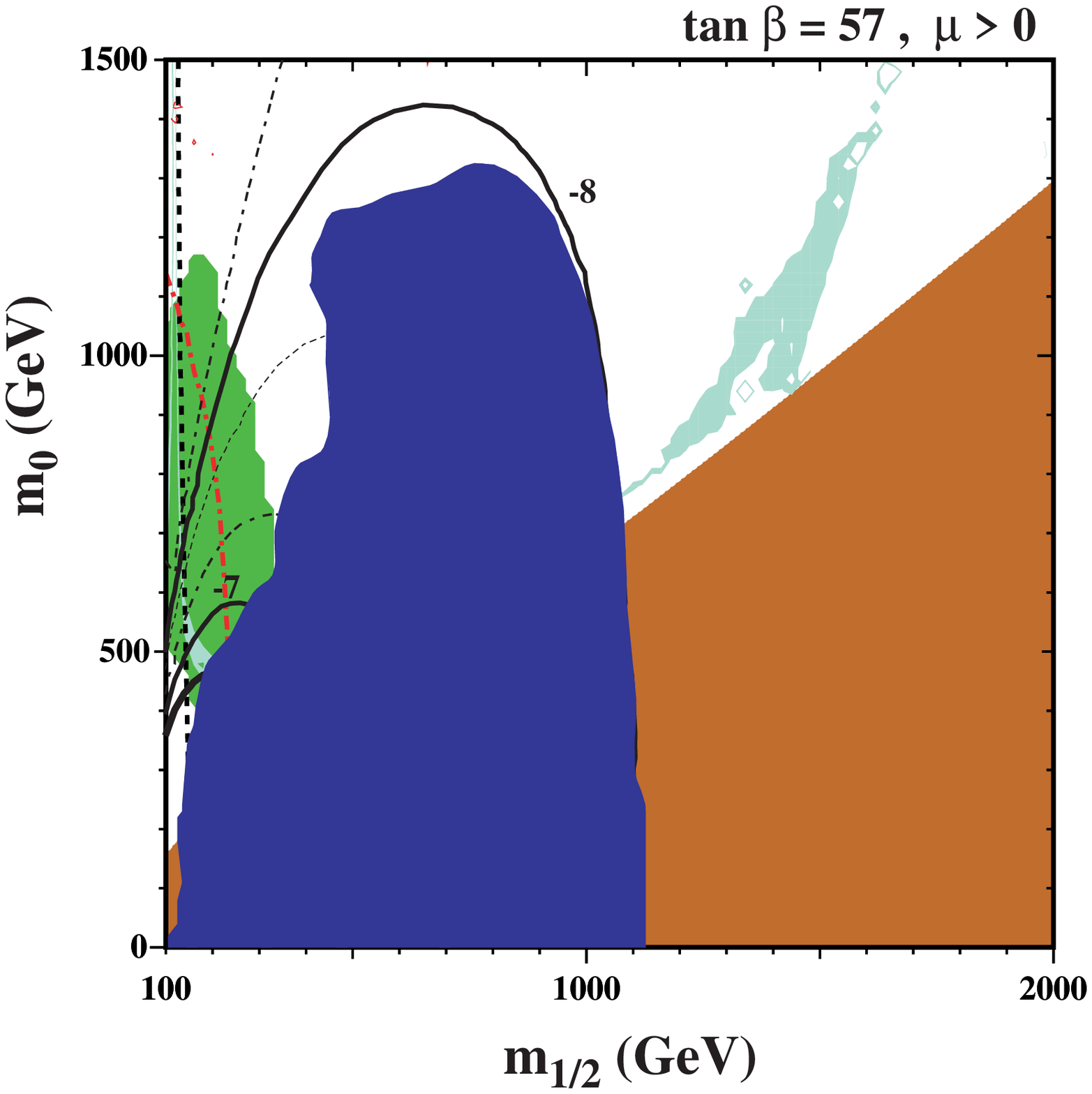,height=3.3in}
\hspace*{-0.17in}
\epsfig{file=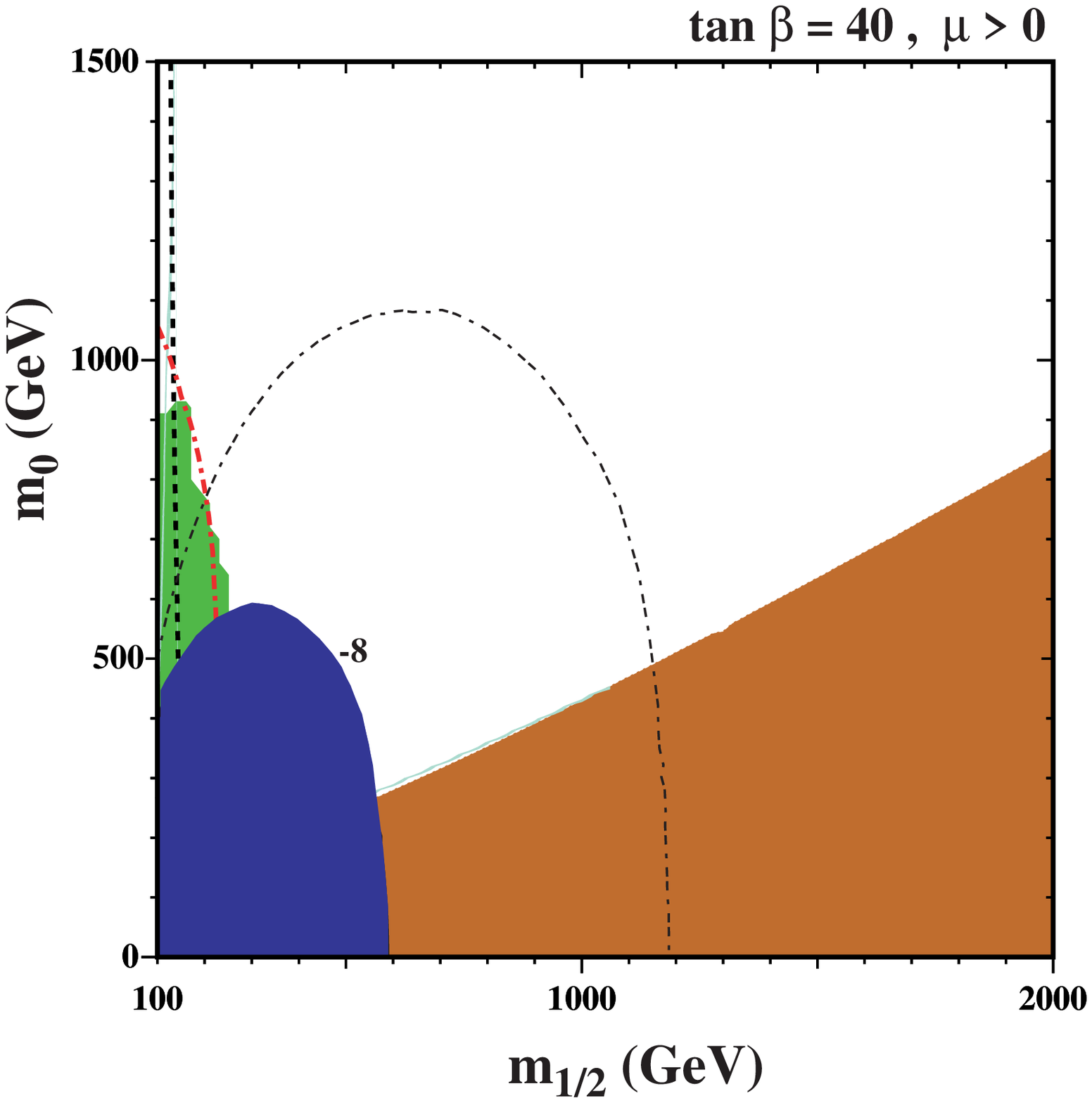,height=3.3in}
\hfill
\end{minipage}
\caption{
{\it
The potential disallowed regions in the $(m_{1/2}, m_0)$ plane for $A_0 = 
0$, $\mu > 0$ and (a) $\tan \beta = 57$, (b) $\tan \beta = 40$, obtained 
assuming a Fermilab Tevatron upper limit on $\bmm$ that is improved 
to a 95\% CL upper limit of $5 \times 10^{-8}$, 
with no parallel reductions in the uncertainties in $f_{B_s}$, 
$m_t$ and $m_b$. Panels (c) and (d) show the corresponding disallowed 
domains assuming a conjectural LHC measurement $BR(\bmm) = (3.9 \pm 
1.3) \times 10^{-9}$. The various contours and shadings in the $(m_{1/2}, 
m_0)$ 
plane are as explained in the text describing Fig.~\ref{fig:CMSSMerrors}.
}}
\label{fig:future}
\end{figure}

It is expected that the Fermilab Tevatron collider experiments will continue
to improve the present sensitivity to $\bmm$ decay.
To assess the likely impact of this improved sensitivity, we
exhibit in Fig.~\ref{fig:future} the potential $(m_{1/2}, m_0)$ planes for
$A_0 = 0$ and $\tan \beta = 57, 40$, obtained using the conservative error
prescription and neglecting possible improvements in the determinations of
$f_{B_s}$, $m_t$ and $m_b$, in the pessimistic case that no signal is
seen.

In panel  (a) of Fig.~\ref{fig:future} for $\tan \beta = 57$, we show  that
the $\bmm$ constraint would, under these pessimistic assumptions, begin to
exclude a region in the neighbourhood of $(m_{1/2}, m_0) = (400, 400)$~GeV
that is favoured by WMAP and allowed by all the other present constraints,
assuming a future 95\% CL upper limit of $5 \times 10^{-8}$.  
The region at small $m_{1/2}$ is allowed because of the high sensitivity to
$m_t$ and $m_b$ seen in panel (b) of Fig.~\ref{fig:errors}.  For $\tan
\beta = 40$, as seen in panel (b) of Fig.~\ref{fig:future}, the region
disallowed by $\bmm$ would still lie within the region already disallowed
by $b \to s \gamma$. The much reduced sensitivity of $\bmm$ decay to $m_t$
and $m_b$ at small $m_{1/2}$ for $\tan \beta = 40$ implies that there is
no allowed `gap' at small $m_{1/2}$.

Panels (c) and (d) of Fig.~\ref{fig:future} show the corresponding
sensitivities at the LHC, assuming a conjectural 3-$\sigma$ measurement
whose central value coincides with the central value predicted by the
Standard Model, i.e., $BR(\bmm) = (3.9 \pm 1.3) \times 10^{-9}$. In
panel (c) for $\tan \beta = 57$ we see that such a measurement would cover
a very large fraction (but not all) of the CMSSM parameter space allowed
by WMAP. On the other hand, the fraction covered in panel (d) for $\tan
\beta = 40$ is somewhat smaller. Of course, it is quite likely that the
present errors in $f_{B_s}$, $m_t$ and $m_b$ will be substantially reduced
by the time of this ultimate LHC measurement. Reducing the $f_{B_s}$
error, in particular, would reduce the scope available for a CMSSM
contribution, and extend the $\bmm$ exclusion region to larger $m_{1/2}$.

\section{Conclusions}

We have seen that the interpretation of the present and prospective
experimental limits on $\bmm$ decay are very sensitive to auxiliary
uncertainties, principally those in $f_{B_s}$, $m_t$ and $m_b$.  At the
present time, these restrict significantly the regions of the CMSSM
parameter space that can be excluded by the current upper limit on this
decay, and their uncertainties may not be reduced significantly during the
remaining operation of the Fermilab Tevatron collider, with the likely
exception of $m_t$. However, one might hope that the uncertainties in each
of $f_{B_s}$, $m_t$ and $m_b$ could be reduced by the time the LHC
achieves its ultimate sensitivity to $\bmm$ decay. As an exercise, we have
considered the possibility that their uncertainties might be reduced to
$\Delta f_{B_s} = 10$~MeV, $\Delta m_t = 1$~GeV and $\Delta m_b =
0.05$~GeV. In this case, the LHC reach in the $(m_{1/2}, m_0)$ plane for
$\tan \beta = 57$ would extend to $m_{1/2} \ga 1400$~GeV along the WMAP
strip.

Beyond the framework of the CMSSM, it would be interesting to study the
interpretation of the $\bmm$ constraint also in the frameworks of more
general models, such as those with non-universal Higgs masses
(NUHM)~\footnote{We expect the situation in the general low-energy
effective supersymmetric theory (LEEST) to be similar.}~\cite{nuhm}. A
complete study of the situation within the NUHM would take us beyond the
scope of this paper, so we restrict ourselves to a few remarks. The models
likely to be disfavoured by $\bmm$ are those with a low value of $m_A$ and
large $\tan \beta$. Such models also tend to predict large
neutralino-nucleus scattering cross sections~\cite{ko}. We have examined a
sample of NUHM scenarios that are apparently excluded by the recent
CDMS~II upper limit on the direct scattering of supersymmetric dark
matter, and found that about half of them are excluded by $\bmm$. The next
step would be to examine models apparently allowed by CDMS~II, to see how
many of them are also excluded by $\bmm$. However, for consistency, one
should also study the effects on the dark matter scattering cross section
of auxiliary uncertainties such as those in $m_t$ and $m_b$, which has not
yet been done in the manner described here for $\bmm$. We plan to present
elsewhere such a unified treatment of the uncertainties.

\section*{Acknowledgments}
\noindent 
The work of K.A.O. and V.C.S. was supported in part
by DOE grant DE--FG02--94ER--40823. We would like to thank 
D. Cronin-Hennessy, C.~Sachrajda 
and M.~Voloshin for helpful discussions.

\end{document}